# Information Theory of Genomes

Dmitri V. Parkhomchuk

Email address: parkhomc@molgen.mpg.de , pdmitri@hotmail.com

**Abstract**

Relation of genome sizes to organisms' complexity is still described rather equivocally. Neither the number of genes (G-value), nor the total amount of DNA (C-value) correlates consistently with phenotype complexity.

Using information theory considerations we developed a model that allows a quantative estimate for the amount of functional information in a genomic sequence. This model easily answers the long-standing question of why GC content is increased in functional regions. The model allows consistent estimate of genome complexities, resolving the major discrepancies of G- and C-values. For related organisms with similarly complex phenotypes, this estimate provides biological insights into their niches complexities. This theoretical framework suggests that biological information can rapidly evolve on demand from environment, mainly in non-coding genomic sequence and explains the role of duplications in the evolution of biological information.

Knowing the approximate amount of functionality in a genomic sequence is useful for many applications such as phylogenetics analyses, in-silico functional elements discovery or prioritising targets for genotyping and sequencing.

**Introduction**

Biological systems can be described on a quite general level as information processing entities. In particular, the carrier of heritable information - DNA in a genome is suitable for consideration in the framework of Shannon's information theory [1]. Information theory describes how information, encoded in discrete variables, is transmitted from sender to receiver over noisy channel with limited capacity (channel's bandwidth). Genomes accumulate and store information about environment and transmit it over generations. The process of information capture from the environment is known as positive or adaptive evolution. However these terms are not invariant in any frame of reference: because the environment is typically heterogeneous in space and time, some information, which is adaptive here and now, may be non adaptive there and then. However after becoming non-adaptive, in some cases, it cannot be easily discarded because the information, which accumulated later, may depend on the context of the previously fixed information, which will be kept for this reason. Thus we refer to the adaptive evolution as the capture of information from the environment. Due to the mentioned environment heterogeneity, the process of capturing, in general, must be continuous. This situation is known as Red Queen concept [2]. By random DNA mutations organisms are effectively trying to guess the information from the environment – where the lucky ones with the correct guesses are granted better survival. This gained information is encoded on DNA and transmitted to progeny. Now the DNA has a limited channel capacity – theoretically at most 2 bits can be encoded per nucleotide, if the frequencies of all 4 nucleotides are equal. It is tacitly assumed that this channel capacity is quite large and is not an issue for biological information – it seems that information density is far from saturation at least in large genomes – due to the observations of spacious "junk" DNA regions, indicating that there is plenty of space with apparently very low information content. However we believe that this assertion is wrong, and in fact, some genomic spots are overloaded with information, where its density is so high that it imposes the pressure on channel capacity, thus forcing the optimisation of DNA content, which maximises the DNA informational bandwidth to the principal limit. Information theory indicates that this bandwidth is at maximum when the entropy of a sequence is at maximum. We propose the following pictorial model of information accumulation in a genome (Fig. 1). For brevity we consider two limiting cases of functional regions – highly functional and "junk" DNA. On the one hand the information density is globally heterogeneous, so there are regions with high and low information density. On the other hand these regions have locally homogeneous information content. Local homogeneity means that near a functional spot there are better chances to find another functional spot. In fact genes' co-clustering is a well-known phenomena. This happens due to the following probabilistic arguments: while random mutations happen equally in both types of regions, the effects are obviously different. In functional regions mutations tautologically are more likely to have functional effects. If this effect is deleterious, the mutation is eliminated from population, while in the case of the "correct guess" the mutation represents some information from environment and thus proceeds further through generations. It is natural to assume that the chances of the *functional* mutation of being advantageous are the same regardless of where it is located - in highly functional or lowly functional genomic region. However the number of functional mutations is much higher in densely functional regions. So we postulate that information accumulation is happening faster in already functional regions (before they are overloaded with information and become too conserved), representing a sort of positive feedback runaway process of information absorption. This almost obvious presumption has an important non-obvious implication – such functional region with continuous information intake should be eventually saturated by the biological information to the point of the DNA maximum informational capacity. The value of maximum content for the biological information depends on its definition and should not be confused with the sequence entropy, though both of them reach maximums in the highly functional regions. The biological information content can be defined by a degree of nucleotide conservation [3], and it does not necessarily reach 2 bits per nucleotide in average because it depends on the encoding methods and other factors. As we will show in humans it is closer to 1 bit in highly functional regions in average, while locally it can be 2 bits for very conserved sequences. However the sequence entropy is well defined and must approach its maximum of 2 bits, or 4 bits in the case of di-nucleotide entropy we used. There can be a subtle source of confusion because in genomics, high functionality implies high conservation, i.e. slowed down sequence evolution, while we posit seemingly the opposite - that highly functional regions are absorbing the environmental information faster than regions with low functionality. However we do not consider some specific functional elements that can be quite highly conserved indeed. We consider large DNA regions with many such elements, so random modifications in the vicinity of these elements (Fig. 1 - appearance of novel binding sites, altering regulation, etc.) often have impact on phenotype and sometimes are advantageous for the current environment. The point is that the later scenario, under reasonable assumptions, is realised more frequently in highly functional regions than in lowly functional regions.

It is worth noting that the initial 'sender' of the information is environment. So it is not the organisms that evolve by themselves to higher complexity. It is the environment 'evolving' (accumulating) to higher complexities, and some organisms, which happened to be exposed to these complexities, may evolve forward for this reason. Of course the environment may get more complex due to the multiple organisms populating and propagating in it, however putting the environment on the top of the information chain (though it is more of a loop) seems to explain the presence of the average direction in evolution (which is not universal for all organisms) more clearly. It is difficult to imagine that a tooth may evolve with no hard food yet available to bite. So the evolution and the presence of such food in environment essentially creates the opportunity - a sort of Platonic form of a tooth, which eventually is guessed out by some organisms.

If we consider the destiny of such highly functional regions then the only radical solution in order to sustain the evolvability further is their duplications. When a region is saturated with the information, its duplication (and subsequent subfunctionalization [4]) may be advantageous because the result of the duplication has the same information content but twice lower functional density, thus higher evolvability (in comparison with saturated sequence) and lower mutation load (which is possible in the presence of adaptation for mutational biases [5]). Noteworthy with the large number of such information 'gates' in a genome, the sexual reproduction is apparently more efficient as these information gates may function in parallel cooperatively (due to recombination) instead of competing with each other for the value of information they pump in [2]. It also indicates the advantage of modularity – where one functional spot is responsible for one function (organ), allowing its independent improvements. In extreme environments (such as high altitudes) where the relative abundance of asexuals was observed, the 'extremity' represents limitations at the principal physical level, thus not constantly supplying novel information, which can be utilised for survival. More information should be gained competing with other organisms in a rich diverse niche than in the one-to-one fight with simple invariant laws of physics. In the later situation the two-fold benefit of asexual reproduction is more important than information assimilation capabilities.

In order to support this general model we have to locate such densely functional regions and to show that they have both - increased purifying selection and increased information accumulation rate (positive selection). Furthermore, the degree of their functionality should comply with the maximum channel capacity requirement – approaching the maximum of sequence variability (entropy). For a genomic sequence alone it seems challenging – to demonstrate the purifying and positive selections simultaneously. However genome-wide SNPs data provides a fairly elegant way to detect both in a set of functional regions without even referring to orthologous comparisons. The purifying selection would lead to decreased SNPs density (as many mutations are deleterious) while the positive selection would lead to increased linkage between SNPs due to a hitchhiking effect during selective sweeps. We used HapMap SNPs database [6] to support this general model and provided few specific organisms examples where the information theory produces coherent estimates of their phenotype and niches complexities while their G- and C-values are discrepant.

**Results and Discussion**

We split human genomic sequence (NCBI build 35) in pieces of 5 kbp, other lengths producing analogous results (Fig. 2). For each 5 kbp genomic piece di-nucleotide entropy was calculated, amount of coding sequences and the residing SNPs from HapMap CEU population. Only the variable SNPs in HapMap CEU [6] population were considered. The di-nucleotide entropy (Fig. 2) was chosen as it represents sequence variability more stringently than single nucleotide entropy, which can reach its maximum in a simple repeat consisting of 'ACGT'. Fig. 3 shows a typical informational "hotspot" containing HOX genes cluster with the exceptionally high sequence entropy indicating intensive information assimilation. Notably the genes in this cluster are the results of duplications.

The linkage was estimated by the mutual information between genotypes of two SNPs $i$ and $j$ (Fig. 4):

$$I_{i,j} = \sum_{g_i=1}^{g_i=3} \sum_{g_j=1}^{g_j=3} P_{ij}(g_i, g_j) \log_2 \frac{P_{ij}(g_i, g_j)}{P_i(g_i) P_j(g_j)},$$

Where $g_i \in \{1,2,3\}$ is the $i$-th SNP genotype, which can take the 3 states – two homozygous and one heterozygous, and $P_i$ are the probabilities of a SNP taking the corresponding genotypes (the probabilities were calculated by occurrences). $I_{i,i}$ in this case is a SNP genotype entropy, which can take values from 0 to 1.5 bits for a SNP at equilibrium. Mutual information as a measure of correlation has certain advantages over other metrics [7].

Besides, the genotype-based linkage calculation is more sensible than haplotype-based because the predicted haplotypes are unavoidably erroneous; furthermore two SNPs association strength in the later case depends (though not strongly) on consideration of SNPs in-between them, which is a statistical oxymoron. Given the genotypes of two SNPs their association strength should not depend on whether we consider the SNPs in-between. Also when selecting the most associated proxy for a SNP ("tagging") the property of mutual information guaranties that the tag SNP with the highest mutual information to the tagged SNP, is the best predictor of its unknown genotype; this simple prediction confidence cannot be guaranteed with other metrics such as $r^2$ and since the observables are the genotypes but not the haplotypes we believe that the mutual information linkage measure is more convenient for many purposes. In any case the trends described here are quite robust and can be reproduced with other measures of linkage because all of them reflect the degree of correlation between SNPs.

Fig. 4 shows how the linkage was calculated. For a given SNP its linkage was described as a sum of its mutual informations to the neighbouring 20 SNPs to the left and to the right (40 in total plus this SNP entropy):

$$L_i = \sum_{j=i-20}^{j=i+20} I_{i,j} \ .$$

The exact number of SNPs does not matter as we tested it with up to few hundreds of neighbouring SNPs. This linkage measure does not take into the account the distance between SNP. So in regions with lower SNP density the linkage is effectively underestimated because with equal correlation strength the pair of SNPs is linked stronger (lower density of recombination events) if the distance between them is larger, however this strengthens our observation of the increased linkage in highly functional regions.

Fig. 5A-C shows the amount of exonic sequence, SNPs density and the amount of linkage versus sequence entropy $H$. Apparently all three properties behave according to the theory – the amount of exonic sequences increase, linkage switches from decrease to increase and SNPs density concordantly decrease. SNPs averaged entropy (Fig. 5D) has the peculiar value of 1 bit probably reflecting the initial population conditions at the recent bottleneck and a very slight increase at functional regions, which is too small to interpret as it can be caused either by positive or by purifying

selection. The initial decrease of linkage is probably due to the increasing with entropy GC content because the recombination rate is proportional to GC content in average. Apparently exons co-localise with the maximum entropy and thus with the GC content of 50%. To the best of our knowledge it is the first theoretical explanation of the long-standing question on why the GC content is increased in functional regions. If exons preferred some different GC content (for example 60%) their maximum density would not coincide with maximum entropy because 60% GC content cannot have the maximum sequence entropy of 4 bits (for 60% the maximum entropy is about ~3.942 bits). For this increased exon density alone, the high-$H$ regions deserved to be called highly functional. However we suggest that the main part of their functionality resides in non-coding sequences. Exons still occupy too small fraction of the high-$H$ sequences (Fig. 5A) to explain the drastic drop of SNPs density (Fig. 5B) and the exons sequences per se cannot contribute significantly to the sequence entropy of their regions.

Since the averaging of parameters was done over large number of genomic pieces falling in the corresponding sequence entropy bins (Fig. 5A), the deviation from mean is rather small, which is also evident from the point-to-point fluctuations (Fig. 5), as each point represent averages over independent sequences; obviously these trends (Fig. 5) are highly significant except for the SNPs average entropy (Fig. 5D).

Armed with this theory we may exercise its predictive power on few known genomes. *D. melanogaster* (fruit fly) and *C. elegans* (nematode) have approximately equal G- and C-values while being of incomparable complexity. So we could expect that the fly has higher information density and maximizes its channel capacity (and hence sequence entropy) and indeed it is obvious from Fig. 6. As we postulated earlier the DNA functional density can be decreased only by duplications (strictly speaking some dramatic environmental changes may also contribute); coincidentally it is well known that fly genome has unusually small numbers of gene duplicates, thus "explaining" its high functional density. Somewhat complementary example are the two fishes – *D. rerio* (zebra fish) and *T. nigroviridis* (puffer fish) which are of comparable complexity but zebra fish has few times larger genome. Again we can see (Fig. 6) that puffer fish has to cope with the unusually small genome by increasing its functional density.

Comparing niches and lifestyles of fruit fly and mosquito we would expect that the later has more information and indeed mosquito genome functional density is similar to that of the fly but it has about three times larger genome size (Table 1). Another pedagogical example is the pair of honeybee (*A. melifera*) and wasp (*N. vitripennis*) (Fig. 6). Although the wasp genome is not available in the assembled form, large amount of whole genome sequencing traces at NCBI is sufficient to estimate its entropy profile. The niche of honeybee is rather static, requiring little information inflow, while more individualistic and predatory wasp faces much more dynamic environment. Fig. 6 shows significant differences in their genome information content – the wasp genome is similar in density to that of mosquito and fruit fly. In this case one can see that the body plan per se takes the small amount of genomic information content while the large part is devoted to environmental interactions.

**Conclusion**

Without referring to the information theory principles it is quite difficult to explain the robust trends of exonic sequence density, SNPs density and the amount of linkage versus sequence entropy (Fig. 5). Given the conceptual simplicity and the predictive power we assume that this model is the most plausible explanation for the observed trends. Or course there could be some other complicating factors involved, for example the linkage may be elevated due to SNPs advantageous covarions being more frequent in functional regions, which would suppress the recombination. Moreover some epigenetic factors like chromatin state or abundance of binding factors may affect the recombination at highly functional regions. Recombination may also be suppressed because some low-complexity recombination-promoting sequences (like ALU) are depleted at highly functional regions. Selective sweep may also cause the decrease in SNPs density, however assuming predominant haplosufficiency of adaptive mutations (as such mutations propagate much faster) and the overall small number of common haplotypes the effect should be rather small in comparison with purifying selection. In either case, the peculiarities of SNPs and exonic densities fit most plausibly in the model of high non-coding functional density at high sequence entropy, while the definite cause of the increased linkage at high-*H* regions requires more careful investigations, because positive selection is difficult to observe genome-wide, in general.

There is an interesting question of why is the entropy systematically lower than 4 bits, as it seems to be a waste of channel capacity. The channel capacity cannot be wasted for no good reason in the saturated functional regions because it would contradict the principles of information theory. The skew of the entropy distributions to the right side (Fig. 6) suggests that there is a sort of barrier at high $H$ resisting further entropy increase. This skew is absent in the genomes, which are far from the barrier such as nematode, zebra fish or yeast (Fig. 6). There could be the interplay of multiple factors – mutation pressure so that some redundancy in information coding is necessary to lessen it, pressure from homogenising action of mutations such as slippage replication, rate of information inflow, frequency of DNA duplications, and the non-random DNA structures involved in epigenetics. By epigenetics we mean not only the heritable information phenomena like methylation, imprinting or histone code, but also the information transmission in somatic cells reproduction – specialised tissues should be differentially programmed epigenetically by chromatin modifications, transcription factors, etc. It is natural to assume that the epigenetic information capacity accumulates in parallel with the genetic one in highly functional regions. Hence the concurrent usage of DNA channel by genetics and epigenetics (when the later 'borrows' some bandwidth from the former by the reserved DNA patterns necessary for binding factors, histones, etc.) makes the DNA variability smaller but the total capacity larger (it can be even more than 2 bits per nucleotide in principle, because the number of epigenetic states can be large) hence the information theory principles are satisfied.

On the other hand the rate of information absorption, which is responsible for the entropy increase may depend on the effective population size and reproduction rate, thus the small population size of mammals (in comparison with insects) may provide the lower drive to increase sequence entropy. In this regard we may note that the effective population sizes of the honeybees (which form the meta-organisms rather than being individual organisms) and more individualistic wasps are likely to be very different, in line with their genomic entropy profiles and the complexities of their niches. However the mouse and human population sizes are likely to differ few times but mouse sequence entropy profile is quite similar to that of human (data not shown).

So we wonder if the larger entropy offset of mammals could be caused by the heavier epigenetics usage due to larger tissues diversity and other factors. On the other side a sequence at maximum entropy stops being evolvable due to very high conservation, and while rather long extremely conserved sequences are known, the fundamental reasons of why they did not reach the maximum sequence entropy are not entirely clear. Also in high-$H$ sequences, we preliminary observed a seemingly paradoxical increase of small-scale (few base-pairs) sequence regularities with the increasing large-scale sequence entropy (and hence functionality), which is in favour of epigenetics-style hypotheses or at least indicates the preference for regular motifs at functional sites. These interesting issues need to be clarified by further investigations.

We can define biological information by the degree of sequence conservation according to [3] where unconstrained sequences have 0 bits and perfectly constrained have 2 bits per nucleotide, so we can plot the biological information against sequence entropy. This definition has probably the fundamental meaning as it coincides in value with the information necessary to locate a binding site in corresponding genomic context [3]. We assume that the information contents of coding, non-coding and epigenetic parts are proportional and the exonic density may serve for the calibration of total genetic information density. So we fit the exonic density with conic section (Fig. 7) (in this case the skewed hyperbola). The theoretical limit for the genetic information content is 2 bits per nucleotide, which may happen at sequence di-nucleotide entropy of 4 bits, thus providing the boundary condition for the fit. The contribution of epigenetics is difficult to define with our limited knowledge of it, so we plotted it for the sake of completeness. Thus we propose the dependence shown on Fig. 7 for the biological information content versus sequence entropy. The relative normalisations may be somewhat species-dependant but the general shape is the same due to the fact that the information theory is quite general and information theoretical properties of genomes are species independent [8]. With these definitions, at highly functional regions, the human sequences reach the maximum of biological genetic information of about 1 bit per nucleotide at $H$~3.97, when averaged over large regions. This value is coherent with the twice-lower SNPs density observed for these regions (Fig. 5). At 2 bits per nucleotide, a sequence cannot accept SNPs by definition. This profile of information density (Fig. 7) apparently explains the behaviour of exonic and SNPs densities (Fig. 5AB). As the information density is known we can estimate information contents of genomes simply by multiplying the density by corresponding sequences lengths. Table 1 shows the information content and density for different species. For some genomes we calculated the density only for a part of genome so the genome length is not shown. As can be seen the G- and C-values large discrepancies are resolved for the pairs of fly - nematode and zebra fish – puffer fish, while in the case of the honeybee – wasp pair, the difference probably reflects the corresponding niches complexities, hence the different information assimilation rates. In contrary to G- or C- values we observe no serious contradictions in the biological information estimates in Table 1 with the corresponding organisms complexities. The most suspicious value is that of mosquito, which is comparable to fishes. Probably the direct comparison of such unrelated species is not entirely legal, because biological information represents essentially algorithmic complexity, as complex patterns can unfold from simple rules. However we suggest that the high value of mosquito reflects the complexity of mosquito niche. Mosquito development cycle is quite complex representing practically different animals (larva, pupa and adult) living in different niches and the adult one feeds on mammals, hence facing much more complex environment in comparison with fruit fly.

Our understanding of non-coding genomic functionality looks rather scarce in comparison with the coding part, which is probably annotated almost completely. With the vast amounts of genomic information already available and much more to come in near future, the practical applications of this theoretical framework seem to be quite numerous. We expect that high-$H$ regions having diverse functional load coincide with biological "regions of interest" as illustrated by HOXA genes (Fig. 3). HOXA genes are intensively studied and were shown to regulate multiple developmental processes. However we found that many genes at high-$H$ regions are of poorly known functions, thus representing the interesting targets for research. We preliminary observed that many of them are transcription factors and are represented widely across phylogenetic tree, as could be expected because the "older" regulatory genes have more time to accumulate diverse complex functions.

Furthermore our results provide another evidence that non-coding DNA represents most of functionality in large genomes and the amount of "junk" DNA should be reconsidered. The increased rate of adaptive evolution (masked by the strong purifying selection) in non-coding regions also explains the difficulties in finding too distant homologies for conserved non-coding sequences, while the homology of corresponding coding regions is still present [9]. Apparently whole-genome genotype-phenotype association efforts could gain efficiency by the improved coverage of high-*H* regions.

**Acknowledgments**

Authors are very grateful to Mikhail Medvedev for the proofreading and valuable suggestions.

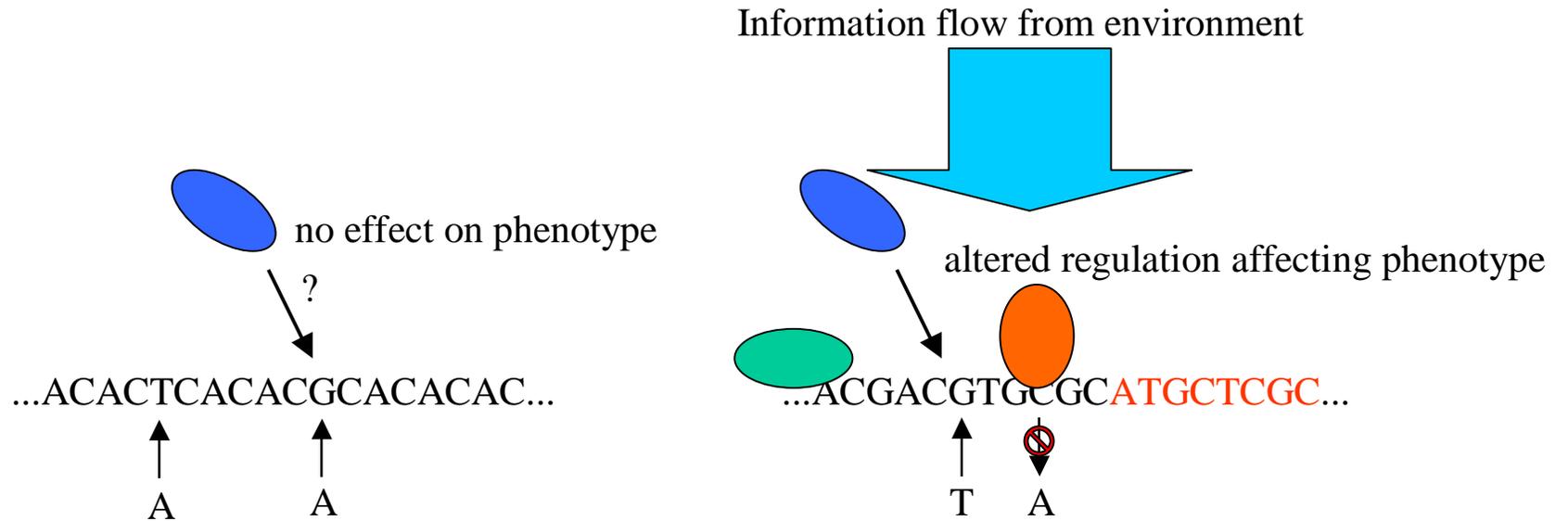

Fig. 1. Highly functional regions although experiencing purifying selection, accept novel information more frequently than regions with low functional density.

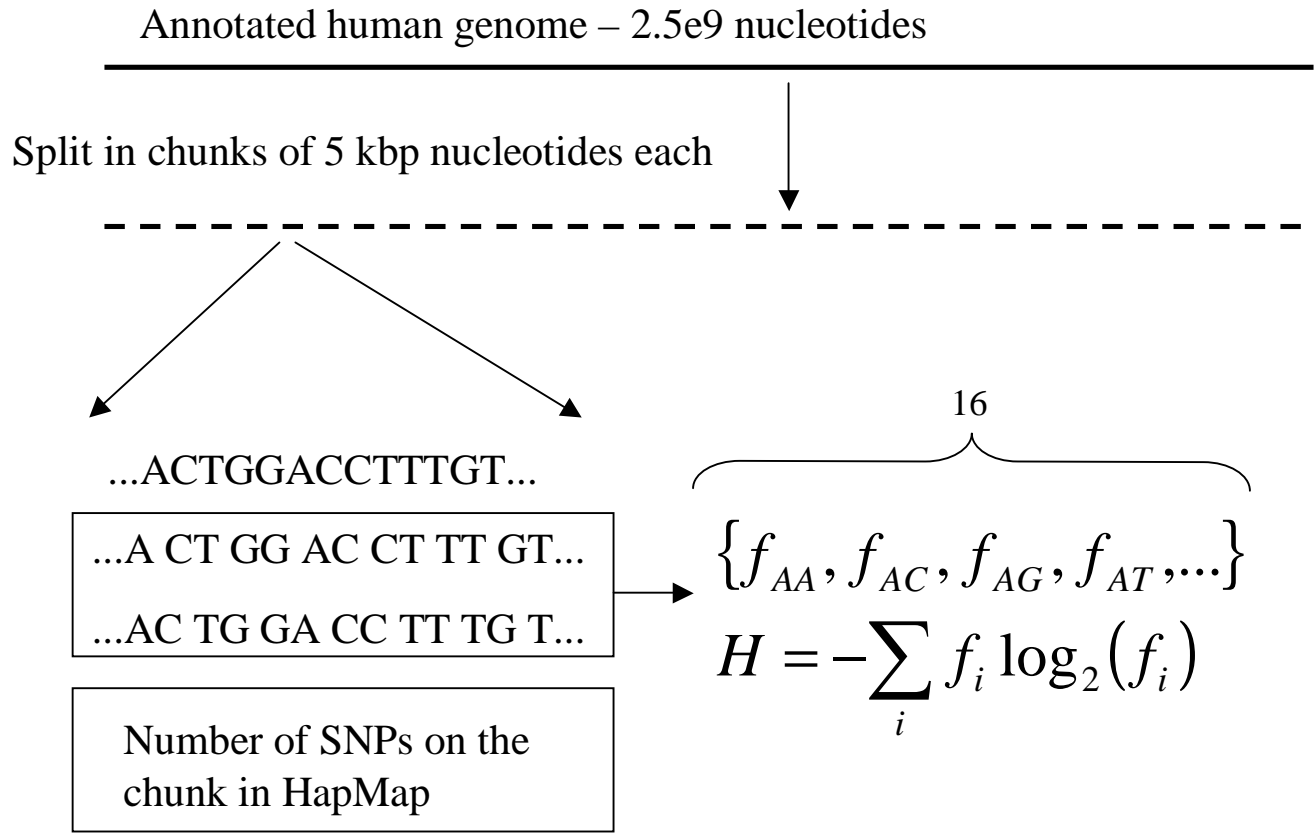

Fig. 2. Schema of data analyses. Besides the sequence entropy for each genomic piece, the number of SNPs in it was extracted from HapMap, and the amount of overlap with known exons.

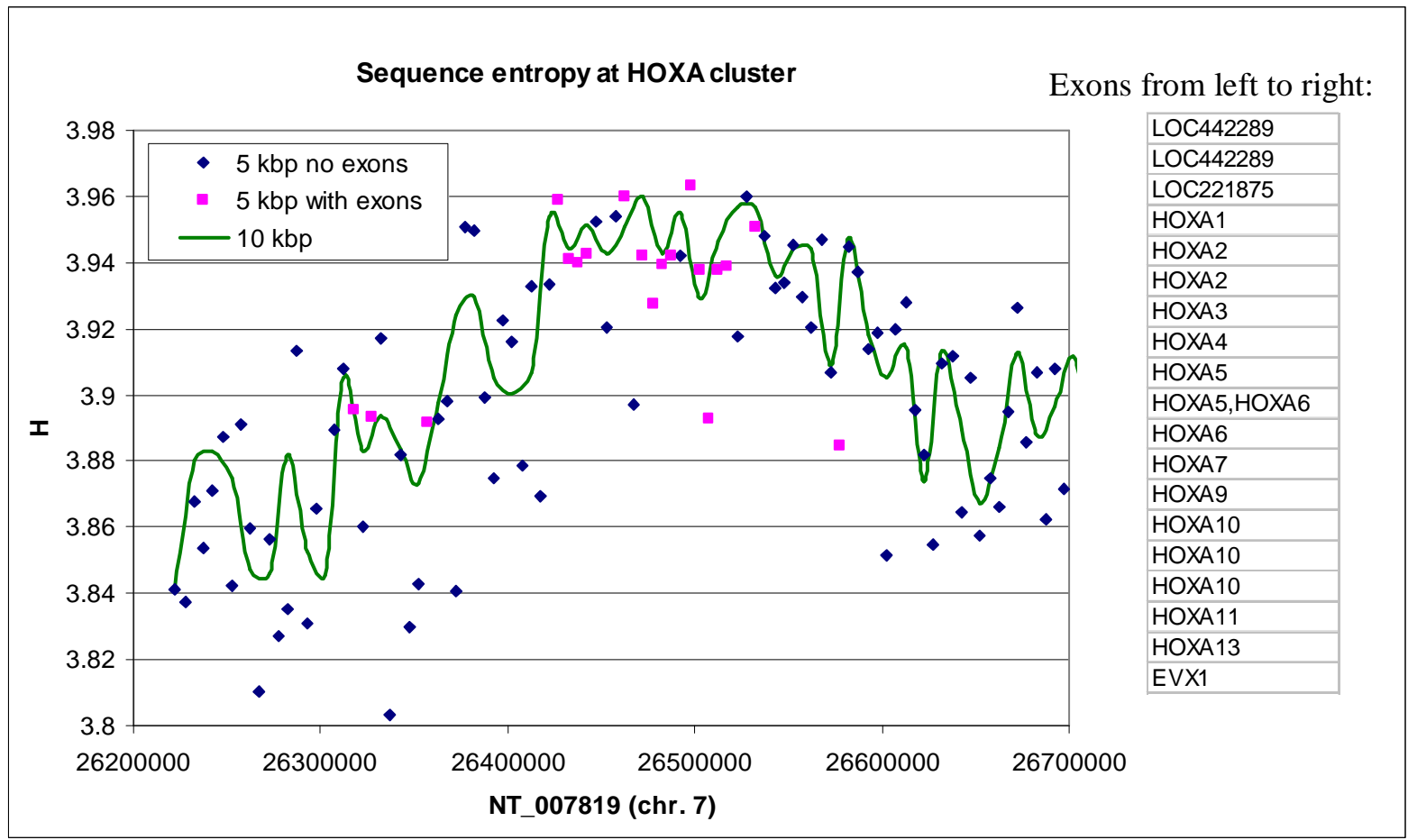

Fig. 3. The entropy of 5 and 10 kbp pieces of the 0.5 mbp genomic sequence containing HOXA genes cluster. The region in the vicinty of the cluster has elevated sequence entropy even in the pieces containing no exons.

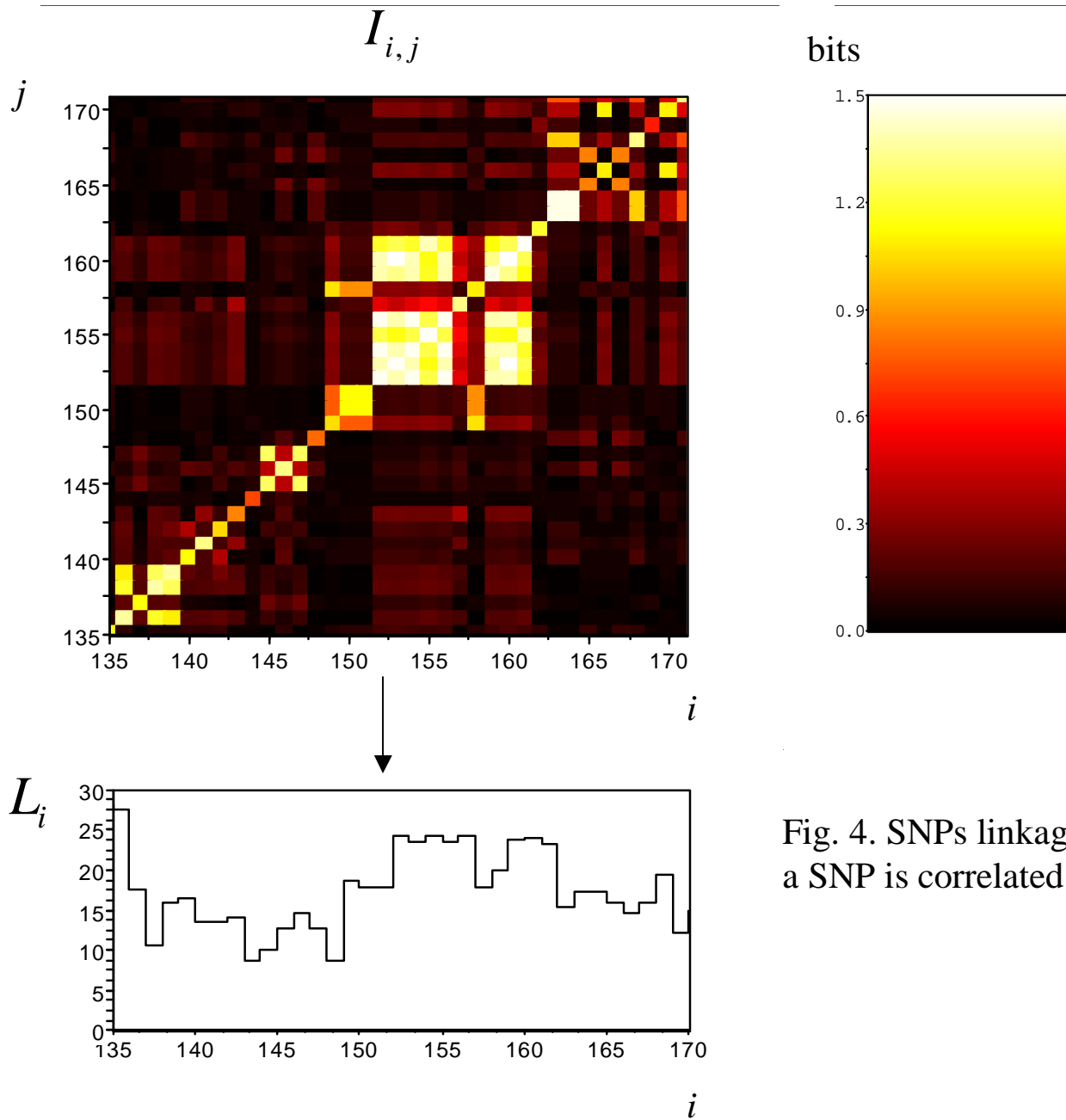

Fig. 4. SNPs linkage reflects how strongly a SNP is correlated to its neighbours.

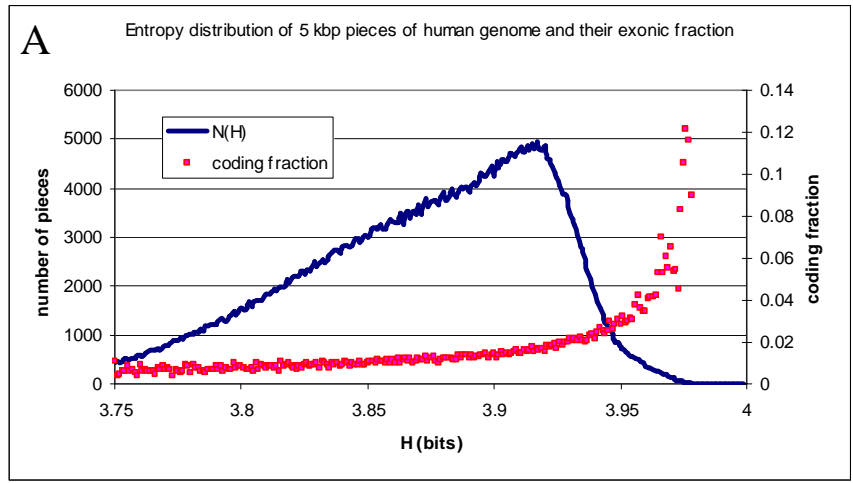
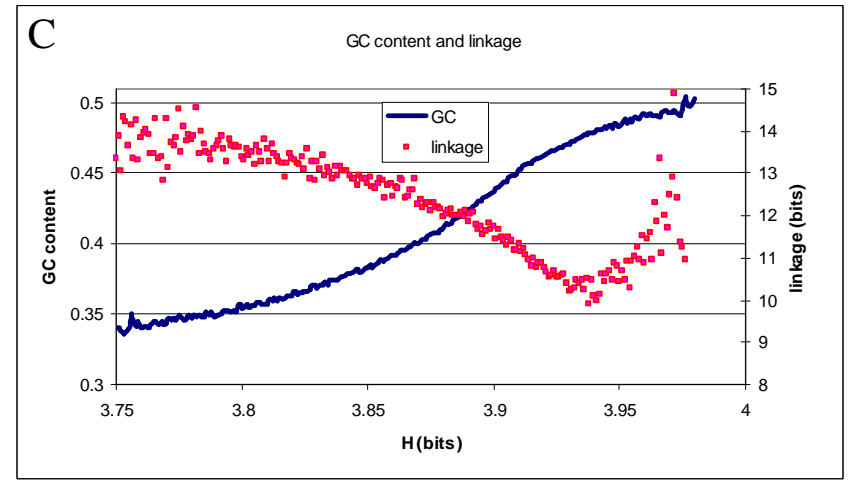
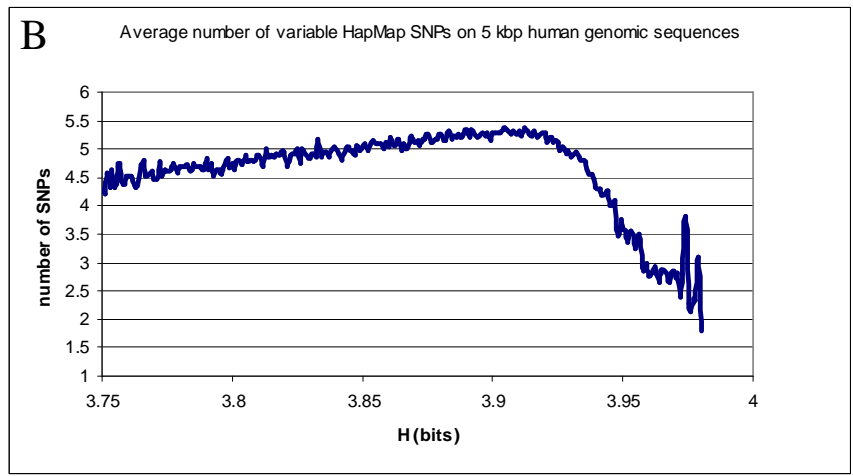
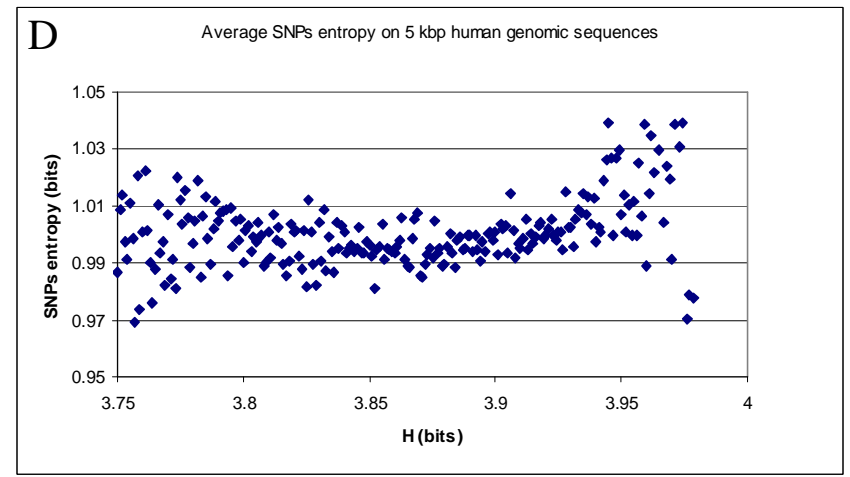

Fig. 5. The averaging of corresponding parametes for 5 kbp genomic sequences was done for N(H) number of pieces, for each entropy bin of size 0.001 bits.

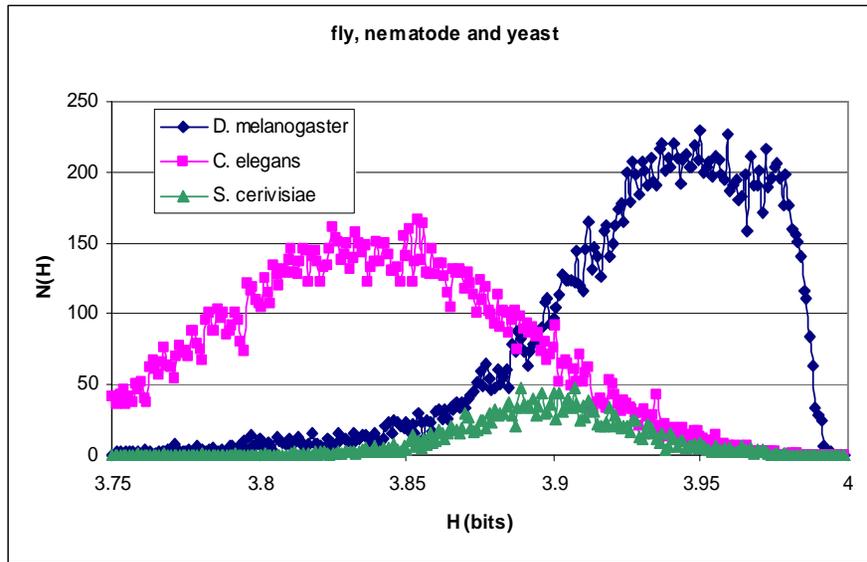
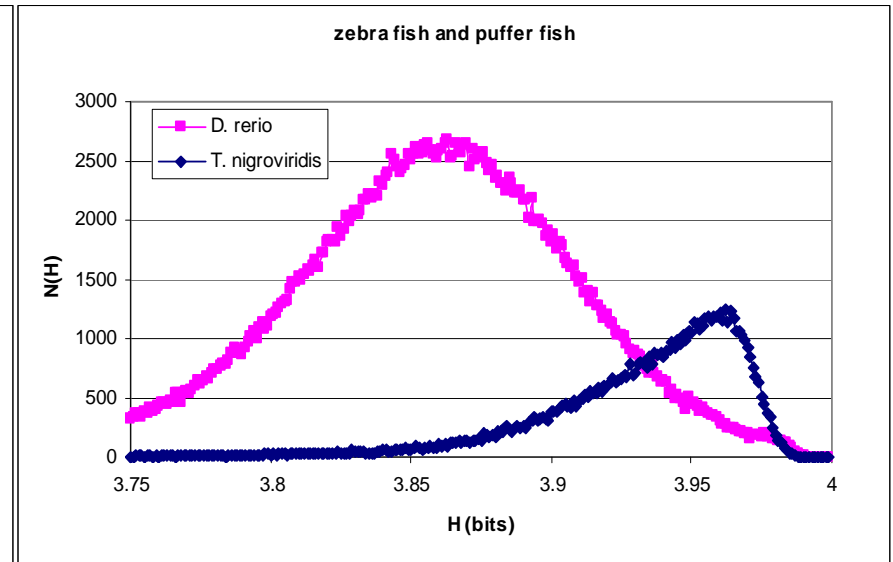
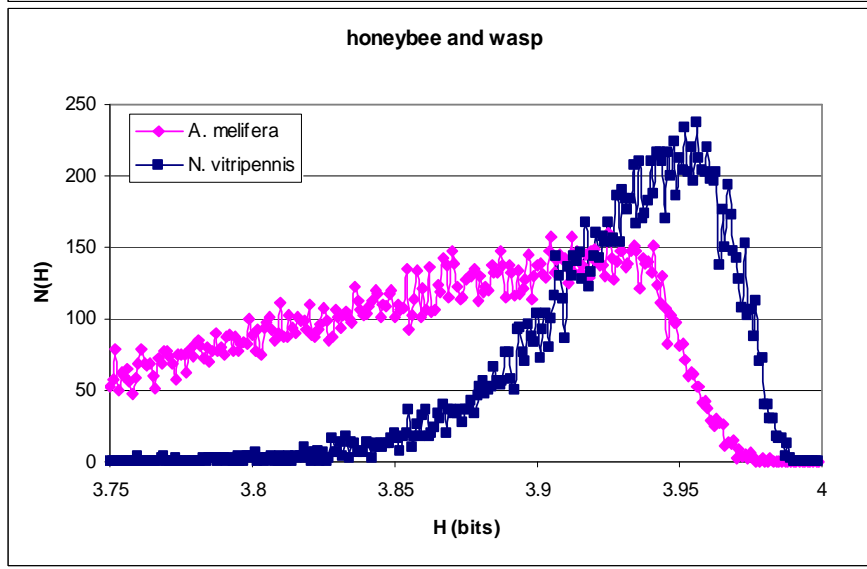

Fig. 6. Dinucleotide entropy profiles for 5 kbp pieces for different genomes.

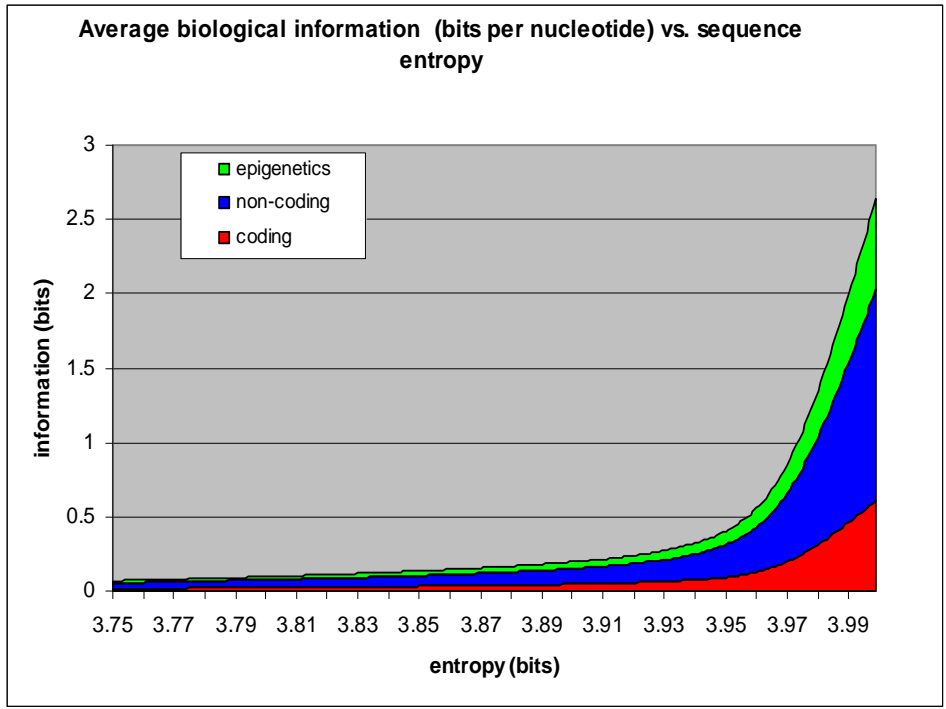

Fig. 7. Biological information density vs. sequence entropy.

| Species | Analyzed genome length (Mbp) | Information density (bits per nucleotide) | Information content (megabits) |
|---|---|---|---|
| D. melanogaster | 95 | 0.21 | 20 |
| C. elegans | 90 | 0.05 | 4.5 |
| H. sapience | 2600 | 0.062 | 160 |
| D. rerio | 1500 | 0.06 | 100 |
| T. nigroviridis | 370 | 0.16 | 60 |
| A. melifera | 110 | 0.063 | 7 |
| N. vitripennis | - | 0.15 | - |
| A. gambiae | 270 | 0.25 | 67 |
| S. cerivisiae | 12 | 0.075 | 1 |
| A. thaliana | - | 0.055 | - |
| G. gallus | - | 0.065 | - |

Table 1. Average information density and content for different species.